\documentclass[journal=nalefd,manuscript=article]{achemso}
\setkeys{acs}{
  etalmode     = truncate,
  maxauthors   = 10
}

\usepackage[version=3]{mhchem} 
\usepackage{xcolor}

\renewcommand\color[2][]{}

\author{Juan~M.~Marmolejo-Tejada}
\affiliation{Department of Chemistry and Biochemistry, Montana State University, Bozeman, MT 59717 USA}
\email{juanmarmolejo@montana.edu}
\altaffiliation{MonArk NSF Quantum Foundry, Montana State University, Bozeman, MT 59717 USA}
\author{Joseph~E.~Roll}
\affiliation{Department of Physics, University of Arkansas, Fayetteville, AR 72701, USA}
\altaffiliation{MonArk NSF Quantum Foundry, University of Arkansas, Fayetteville, AR 72701, USA}
\author{Shiva~Prasad~Poudel}
\affiliation{Department of Physics, University of Arkansas, Fayetteville, AR 72701, USA}
\altaffiliation{MonArk NSF Quantum Foundry, University of Arkansas, Fayetteville, AR 72701, USA}
\author{Salvador~Barraza-Lopez}
\affiliation{Department of Physics, University of Arkansas, Fayetteville, AR 72701, USA}
\altaffiliation{MonArk NSF Quantum Foundry, University of Arkansas, Fayetteville, AR 72701, USA}
\email{sbarraza@uark.edu}
\author{Mart\'in~A.~Mosquera}
\affiliation{Department of Chemistry and Biochemistry, Montana State University, Bozeman, MT 59717 USA}
\altaffiliation{MonArk NSF Quantum Foundry, Montana State University, Bozeman, MT 59717 USA}
\email{martinmosquera@montana.edu}

\title{Slippery paraelectric transition metal dichalcogenide bilayers}

{\color{blue}
}

\begin{document}

\begin{tocentry}




\centering
\includegraphics[width=1.0\textwidth]{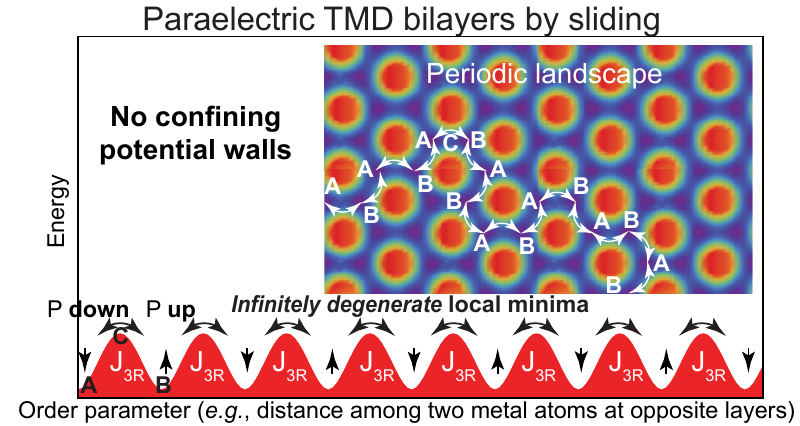}
\end{tocentry}

\begin{abstract}
Traditional ferroelectrics undergo thermally-induced phase transitions whereby their structural symmetry increases. The associated higher-symmetry structure is dubbed {\em paraelectric}. Ferroelectric transition metal dichalcogenide bilayers have been recently shown to become paraelectric, but not much has been said of the atomistic configuration of such a phase. As discovered through numerical calculations that include molecular dynamics here, their paraelectricity can only be ascribed to a time average of ferroelectric phases with opposing intrinsic polarizations, whose switching requires macroscopically large areas to slip in unison.
\end{abstract}



Ferroelectrics are ubiquitous within capacitors\cite{Susan}, and could also be of use on neuromorphic computers.\cite{neuromorphic} Ferroelectricity arises on  {\color{black}materials} lacking a center of inversion and layered materials offer two novel pathways to induce ferroelectricity:
(a) by thinning down,\cite{Kai2016,review,ferroelectricmetal} or
(b) by applying a relative rotation and/or sliding \cite{hBN0,hBN1,hBN2,Liu,pablo}.
Sustained experimental and theoretical efforts to characterize layered ferroelectrics focus on (i) the atomistic nature of the ferroelectric-to-paraelectric two-dimensional phase transition, and (ii) the deployment of critical temperature ($T_C$) trends.

And thus, the ferroelectic to paraelectric phase transition for a family of ferroelectrics created by the thinning down of their layered bulk and known as group-IV monochalcogenide monolayers (IVMMLs) \cite{MehboudiNL,wu,Kai2016,qian} is facilitated by the rotation of individual metal-chalcogen dimers within a given monolayer \cite{Villanova} taking place within a sub-picosecond timeframe \cite{MehboudiPRL,MehboudiNL,shiva,KaiNL}.

{\color{black}On the other hand}, experimental work on ferroelectrics created by the relative rotation and/or sliding of two monolayers \cite{Franchini,acsnano.7b02756,2018paper,park2019ferroelectric,Wu21,new} (type (b) ferroelectrics) is relatively newer \cite{Liu,pablo}, but  Liu and coworkers have demonstrated a transition from a ferroelectric configuration (one in which the intrinsic electric dipole moment $P$ is finite) onto a paraelectric one (in which $P=0$) at finite temperature {\color{black}unequivocally}, {\color{black} creating} electronic devices based on $3R$ transition metal dichalcogenide bilayers (TMDBs) \cite{Ciraci,avouris_heinz_low_2017} for this purpose \cite{Liu}.

What must atoms do to turn those bilayers from ferroelectric ($P\ne 0$) to paraelectric ($P=0$)? In other words, what is the atomistic structure of the experimentally verified paraelectric phase of TMDBs \cite{Liu}? The mechanism, unveiled here, {\color{black} turns out to be a realization of brownian motion on a honeycomb lattice}.\cite{Pradeep1}

We start reviewing the two stacking configurations for the TMDB, known as $2H$, or $3R$. The discussion includes energy landscapes and a comparison of global and local energy minima. Molecular dynamics (MD) calculations with force fields designed via machine-learning are deployed for four TMDBs at multiple temperatures {\color{black}afterwards}. Those calculations reveal that the ferroelectric to paraelectric two-dimensional structural transformation is facilitated by  sliding {\color{black}events on the honeycomb lattice, with polarization $P$ changing direction at every discrete step.} $T_C$ is shown to be proportional to predetermined energy barriers. Conclusions are provided at the end.


MoS$_2$, WS$_2$, MoSe$_2$, and WSe$_2$ bilayers were studied. Pending a detailed description of  Methods, this computational study was divided in two subcategories:  {\color{black}(i)} {\em zero-temperature} density functional theory (DFT) calculations that utilize a plane wave set were performed to determine relative structural energies, as well as the vibrational properties of transition metal dichalcogenide bilayers, {\color{black}and (ii)}  {\em finite temperature} MD calculations that relied on a DFT numerical atomic orbital basis set for efficiency, from which machine-learned classical interatomic potentials were obtained.

The {\color{black}studied} materials were ordered by their {\em mean atomic number} $\bar{Z}$, defined as $1/3\sum_i Z_i$, where $Z_i$ is the atomic number of any of the three atoms in a given monolayer unit cell. Table \ref{ta:Ta1} displays $\bar{Z}$ and the energy  {\color{black}cost} $\Delta E$  {\color{black} to modify} the bilayer from a $2H$ configuration onto the $3R$ one. $\Delta E$ increases with $\bar{Z}$. The {\color{black}rotation} process can be found as Supporting Information (SI).

When comparing {\color{black}our} results {\color{black}against} experimental ones, one must remember that the $3R$ rotated bilayer is buried within a bulk sample in the latter case: a TMD sample is cleaved by a shear strain that induces a relative rotation of the two cleaved parts, which remain bound after the mechanical manipulation. In this study, on the other hand, a bilayer exposed to vacuum on both ends is being considered. We are unaware of the experimental observation of spontaneous rotations of a $3R$ bilayer system back onto the $2H$ ground state: the mechanical energy utilized to rotate exceeds any thermally activated barrier to undergo a macroscopic in-sync rotation of half a layered material. Here, periodic boundary conditions preclude the two layers from undergoing relative rotations.

\begin{figure*}
\begin{center}
\includegraphics[width=0.96\textwidth]{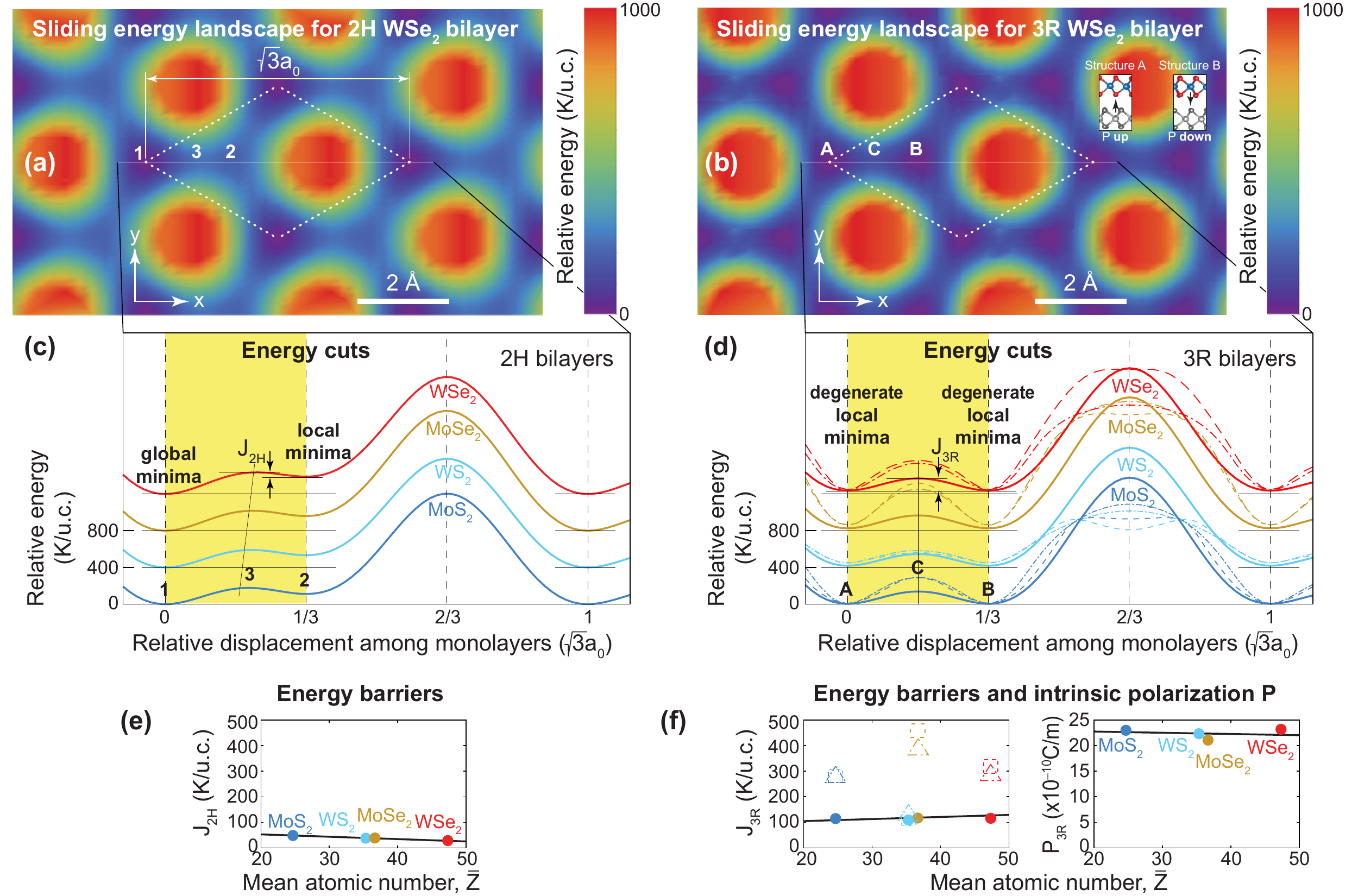}
\end{center}
\caption{Energy landscape for (a) $2H$ and (b) $3R$ {\color{black}WSe$_2$ bilayers} as a function of the relative sliding among {\color{black}their constituent} monolayers. {\color{black} (c) and (d): C}uts along a horizontal line on the landscapes, in which local and global minima, as well as energy barriers ($J_{2H}$ and $J_{3R}$) for multiple TMDBs are displayed. Solid lines were obtained with a plane-wave DFT method, dashed lines on plot (d) are obtained with a DFT method using a localized basis set, and dash-dot lines {\color{black}on (d)} indicate the energy landscape as obtained from a classical force field relying on machine-learning techniques. Relative energy differences among the $2H$ ground state and the degenerate $3R$ minima are consistent with Table \ref{ta:Ta1}. {\color{black} (e) and (f): E}nergy barriers among local minima and the smallest energy barrier, and $P$ for the local minima in the $3R$ phase. $P$ flips sign in going from one local minima to the nearest one.\label{fig:Fig1}}
\end{figure*}

Figure \ref{fig:Fig1}(a) shows the energy that it takes for two layers in the $2H$ WSe$_2$ bilayer to slide with respect to one another, while a similar plot on {\color{black}Figure} \ref{fig:Fig1}(b) corresponds to the $3R$ WSe$_2$ bilayer. The  multiple (purple) minima on Figure \ref{fig:Fig1}(a) encode ground state, centrosymmetric $2H$ bilayer configurations, and the dashed diamond {\color{black}indicates} the  {\color{black}area} of a unit cell. The $2H$ bilayer also displays a shallow local minima, {\color{black}seen} on a light blue  {\color{black}color on Figure \ref{fig:Fig1}(a)}. The distance from the global minima to the nearest local minima is $a_0/\sqrt{3}$, with $a_0$ the bilayer's lattice constant.

Figure \ref{fig:Fig1}{\color{black}(c)} is a one-dimensional cut of the energy landscape, taken along the white horizontal line {\color{black}displayed} on {\color{black}Figure \ref{fig:Fig1}(a)}. Scaling the  {\color{black}horizontal axis} by their respective {\color{black}$a_0$}, the plot includes {\color{black}energetics for the} MoS$_2$, WS$_2$, and MoSe$_2$ $2H$ bilayers; each plot was vertically displaced by 400 K/u.c.~for an easier comparison.

There is a tall energy barrier located at about 2/3 of $\sqrt{3}a_0$ {\color{black}on Figure \ref{fig:Fig1}(c)}. Each subplot {\color{black}on Figure \ref{fig:Fig1}(c)} also displays a global minima labeled $\mathbf{1}$, and a local minima labeled $\mathbf{2}$. In between these minima, there is a smaller energy barrier $J_{2H}$ that decreases with $\bar{Z}$ {\color{black}(Figure \ref{fig:Fig1}(e))}.

Figure \ref{fig:Fig1}(b) highlights the {\color{black}energy landscape of the $3R$ WSe$_2$ bilayer, which contains} {\em periodically-spaced and degenerate}  local {\em mimima} {\color{black}(see Table \ref{ta:Ta1})}. The two degenerate minima are labeled $\mathbf{A}$ and $\mathbf{B}$, and the point at the height of the local barrier is dubbed $\mathbf{C}$ in that middle subplot \cite{MehboudiNL,MehboudiPRL,prb2018,tyler,review}. Similarly, the energy barrier was called $J_{3R}$, and each trace was displaced by 400 K/u.c.~for {\color{black}easy comparison}. Unlike ferroelectric IVMMLs that have orders-of-magnitude tunability of their energy barriers with chemical composition \cite{MehboudiNL}, $J_{3R}$ remains order-of-magnitude similar for all studied compounds here. Its magnitude--above 100 K/u.c.--precludes quantum tunneling among wells \cite{tyler,PhysRevB.104.L060103}, so that {\color{black}traversing} from one local minima to the nearest one {\color{black}can be understood as} a classical process.

Both local minima $\mathbf{A}$ and $\mathbf{B}$ {\color{black}on Figure \ref{fig:Fig1}(b)} lead to a non-centro-symmetric bilayer structure with a net electric dipole moment $P$ \cite{Franchini,acsnano.7b02756,2018paper,park2019ferroelectric,Wu21,new} (Figure \ref{fig:Fig1}{\color{black}(f)}). Unlike IVMMLs which have an in-plane $P$ \cite{MehboudiPRL,Kai2016,qian,KaiNL} ({\em i.e.}, $P$ pointing along their periodic direction), the {$3R$} bilayers have an out-of-plane $P$ {\color{black}when at their local minima configurations (see insets on Figure \ref{fig:Fig1}(b))}. The Berry phase approach for intrinsic polarization \cite{vanderbilt} was applied on a periodic bulk bilayer configuration, and its value multiplied by the lattice constant along the direction parallel to $P$ to report two-dimensional values.

While the contents of Figure \ref{fig:Fig1} were obtained with a DFT tool that utilizes plane waves to expand electronic states \cite{vasp-1,vasp-2,vasp-4} and employed exchange-correlation potentials to describe crucial van der Waals forces without empirical fitting parameters,\cite{Klime_2009,Klime_2011} this approach is prohibitive for MD calculations with trajectories spanning a micro-second, and we use machine-learning-based interatomic forces based on moment tensor potentials (MTPs) for that purpose \cite{Shapeev2016,Mortazavi2020,Mortazavi2020a,Rosenbrock2021,Gubaev2019,Nyshadham2019,Juan1}. Those are based on a different DFT tool \cite{atk,Smidstrup2020} that includes van der Waals interactions semi-empirically \cite{Grimme2006}, and whose results are now discussed.

Dashed curves {\color{black}on} Figure \ref{fig:Fig1}{\color{black}(d)} were obtained with the DFT code from which machine-learning force fields are obtained, while dash-dot curves are {\color{black}obtained using the machine-learned} force fields. Discrepancies among these two curves are minimal around the small barrier $J_{3R}$. On the other hand, careful analyses have shown a co-dependency of energy barriers on exchange-correlation potentials \cite{shiva,review} and on the DFT code employed. In that sense, the values of $T_C$ to be reported here should not be considered quantitative predictions for experiment, but order-of-magnitude correct only\cite{review}. The observed phenomenology is the important contribution here.

\begin{figure}
\begin{center}
\includegraphics[width=0.48\textwidth]{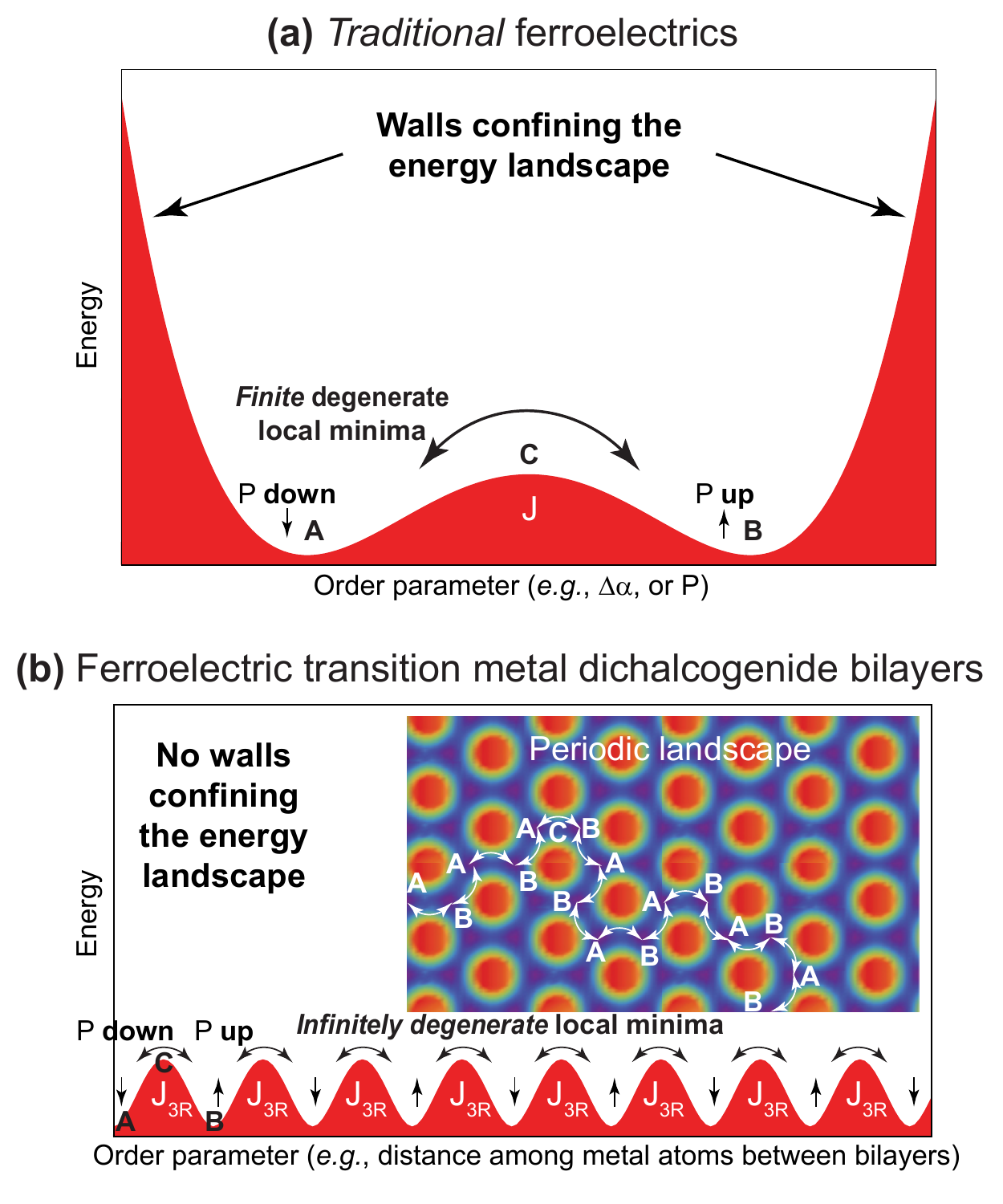}
\end{center}
\caption{(a) Ferroelectrics {\color{black}are traditionally} described by a polynomial energy landscape with {\color{black}two} degenerate minima, an energy barrier $J$, and {\color{black}energy-}confining walls. (b) $3R$ transition metal dichalcogenide bilayers furnish an ``unusual'' ferroelectric with an infinite number of degenerate minima {\em on a periodic energy landscape}. A ``paraelectric'' state is the time-average of $P$ over long times--in which $P$ takes definite non-zero values that swap sign at any given time, averaging down to zero.\label{fig:Fig2}}
\end{figure}

Figure \ref{fig:Fig2} was designed to posit an unusual hypothesis within ferroelectrics: {\color{black}Indeed,} Figure \ref{fig:Fig2}(a) displays a ``common'' energy landscape for ferroelectrics having two energy minima. In most cases, a structural order parameter such as a distance or an angle $\Delta \alpha$ \cite{Kai2016,prb2018} can be linearly linked to $P$, so that the horizontal axis can be thought of representing either the structural order parameter of $P$ interchangeably\cite{MehboudiNL,prb2018,review}. This energy landscape is {\em aperiodic}: there are tall confining energy walls, constraining the order parameter from moving too far away from the local minima. As a result, average quantities computed on the landscape {\color{black}coalesce to definite values} as the energy barrier $J$ is overcome. (Considering $\Delta \alpha$, its value turns to zero at a certain $T_C$.)

{\color{black}On the other hand, and quite distinctly,} the energy landscape on Figure \ref{fig:Fig1}(b){\color{black}--reproduced over a larger spatial region on Figure \ref{fig:Fig2}(b)--}displays a periodically-placed, macroscopically large number of energy minima (an {\em infinite number} for an ideal crystal). This must be so because, after all, transition metal dichalcogenides are dry lubricants. This picture is markedly different from the one presented {\color{black}on Figure \ref{fig:Fig2}(a) and employed} in Refs.~\cite{Liu} and \cite{new}, in which a double-well energy functional is still being considered and for which a single, {\em definite} paraelectric atomistic configuration can be created. {\em Such realization is the main point of this Letter.} Indeed, the emerging picture for paraelectric phenomena on TMDBs (Figure \ref{fig:Fig2}(b)) is one in which temperature rises sufficiently enough such that the barrier $J_{3R}$ can be traversed by all atoms on a macroscopic monolayer; something that might be statistically rare, and unlike anything seen before {\color{black} within the fields of two-dimensional phase transtions}\cite{qian2} {\color{black} and ferroelectrics\cite{rabe2010physics}}. There is a plethora of possible local minima structures to jump from/to, as opposed to just two, which would be the case on a Landau theory.

The picture presented on Figure \ref{fig:Fig2}(b){\color{black}--in which monolayers slide in discrete steps along a honeycomb lattice--}is true: working with MoS$_2$ bilayers, it was demonstrated that a shear phonon {\color{black}mode} can be activated to change polarization \cite{park2019ferroelectric}. It is shown here that shear {\color{black}can be} thermally activated to effect paraelectric behavior \cite{Liu}, and the responsible vibration mode is shown as SI.

The trained MTPs were used to simulate a $3R$ WSe$_2$ bilayer for up to one microsecond of MD evolution with a 10 femtosecond time step; a runtime orders of magnitude larger than those reported for other 2D ferroelectrics before \cite{MehboudiNL,MehboudiPRL,prb2018}.
In more detail, classical MD calculations employing the NVT ensemble (one in which the number of atoms, containing volume, and target temperature are kept fixed) were performed on a 5$\times$5$\times$1 supercell that contains 150 atoms. The target temperature was set with a Nos{\'e}-Hoover thermostat. The use of an NVT ensemble as opposed to the NPT ensemble \cite{MehboudiNL,MehboudiPRL,prb2018,tyler,review} is due to the fact that TMDs are sturdier than other 2D materials that undergo rectangular-to-square phase transformations~\cite{shiva} and no significant in-plane compression is to be expected.

Prior experience indicates a relation among an energy barrier $J$ and $T_C$ of the form $T_C \simeq 1.5J$, when $J$ is expressed in K/u.c.~\cite{prb2018,tyler,review}. The moment tensor potential (MTP) value for $J_{3R}$ on Figure \ref{fig:Fig1}(b) turned out to be 319 K/u.c., which suggests a $T_C$ near 478 K. To verify such hypothesis, 1 $\mu$s calculations were ran at ten distinct target temperatures (100, 200, 300, 400, 460, 480, 490, 500, 510, and 530 K).

\begin{figure*}
\begin{center}
\includegraphics[width=0.96\textwidth]{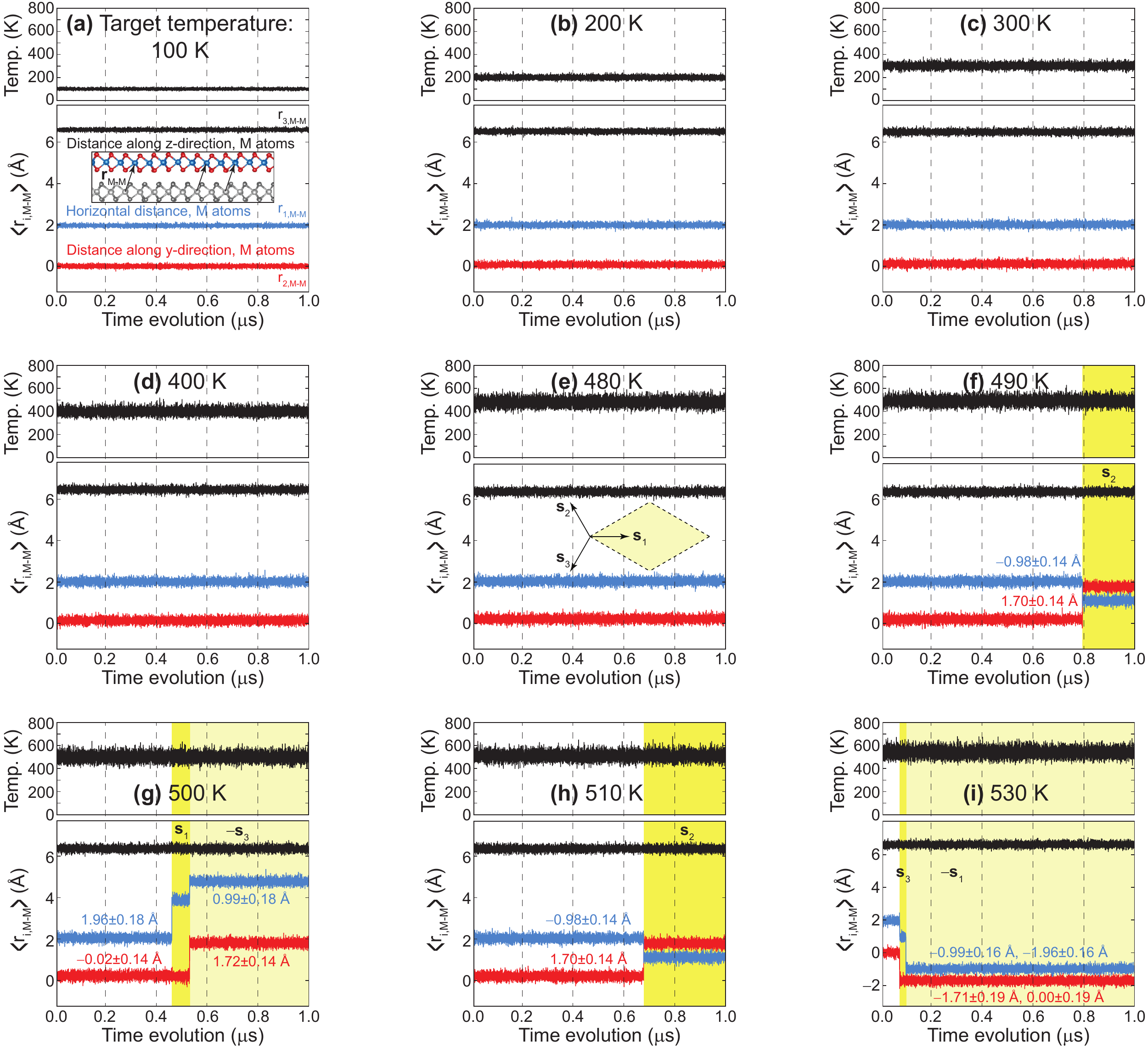}
\end{center}
\caption{{\color{black}Demonstrating the temperature-activated relative sliding} of the $3R$ WSe$_2$ bilayer: $\mathbf{r}_{M-M}=(r_{1,M-M}, r_{2,M-M}, r_{3,M-M})$ is the vector joining pairs of W atoms {\color{black}belonging to opposite monolayers} on {\color{black}the same} unit cell (insets on subplot (a)), and its average over 25 unit cells per frame is tracked {\color{black}as a function of time by blue, red, and black traces}; $r_{1,M-M}=0.00$ \AA, $r_{2,M-M}=1.96$ \AA, and $r_{3,M-M}=6.59$ \AA {} at zero temperature. Sliding events are {\color{black}observed}  on subplots (f) through (i); the magnitude of those displacements is consistent with the vectors drawn as an inset in subplot (e) {\color{black}which furnish a honeycomb lattice}. We assign a critical temperature $T_C$ to the temperature for which the first sliding event occurs within the full one microsecond simulation time, understanding that $P$ will become zero as a long-time-average. $T_C$ as extracted from these plots has a $\pm$ 10 K resolution. The sliding events seen here validate the hypothesis raised on Figure \ref{fig:Fig2}(b).\label{fig:Fig3}}
\end{figure*}

\begin{figure}
\begin{center}
\includegraphics[width=0.48\textwidth]{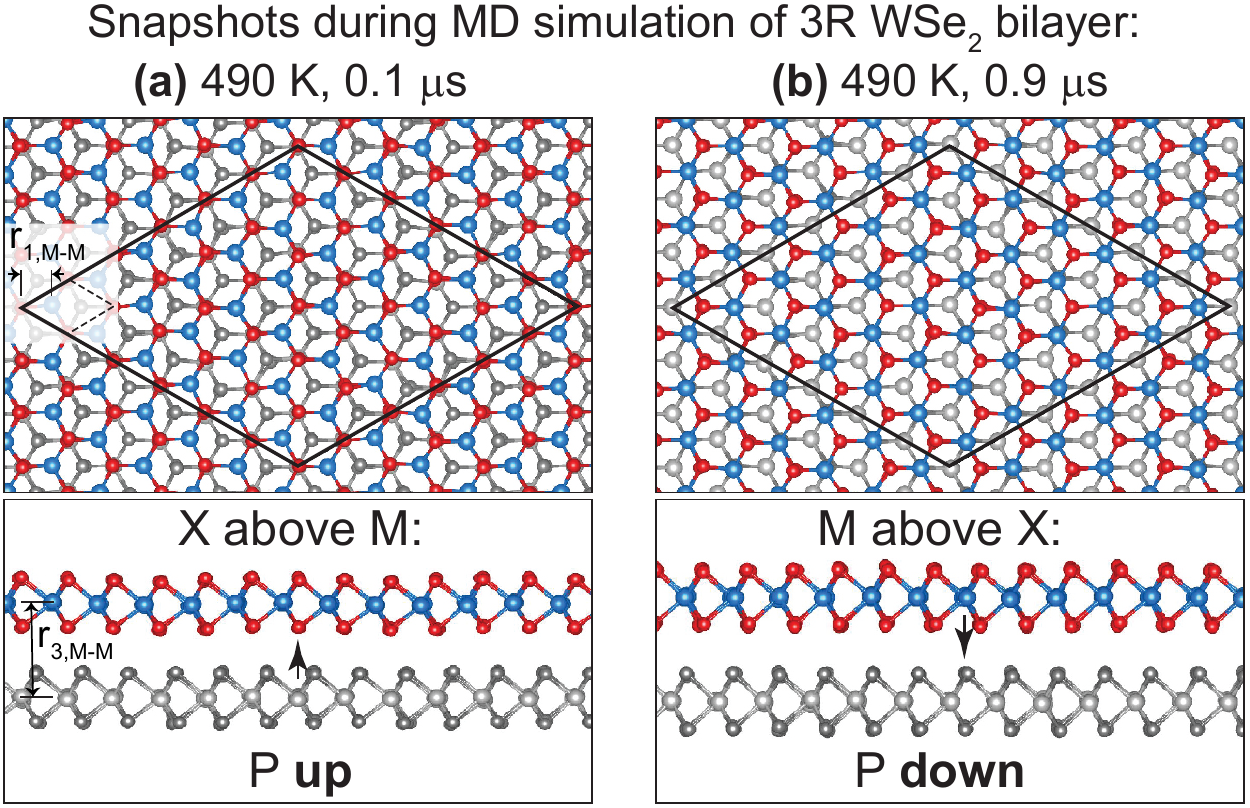}
\end{center}
\caption{Sliding $3R$ WSe$_2$ bilayer at 490 K--in which $P$ swaps sign--as seen from two MD snapshots. {\color{black}The $x-$ and $y-$components of $\mathbf{r}_{M-M}$ are highlighted on a unit cell ($r_{2,M-M}$ is nearly zero and not shown for that reason)}.\label{fig:Fig4}}
\end{figure}

At each MD frame, we tracked the instantaneous temperature $T$ and the average separation $\langle \mathbf{r}_{M-M}\rangle=(\langle r_{1,M-M}\rangle, \langle r_{2,M-M}\rangle,\langle r_{3,M-M}\rangle)$ among the two closest metal atoms at each unit cell, out of the 25 individual unit cells that are available at each frame. Three such vectors are schematically shown at an inset on Figure \ref{fig:Fig3}(a){\color{black}, and projections onto the $x$ and $z-$ axes can be found on Figure \ref{fig:Fig4}(a)}. When temperatures are in between 100 and 480 K (subplots \ref{fig:Fig3}(a) through \ref{fig:Fig3}(e)), distances among metal atoms remain on track, with fluctuations of the order of $\sim$0.14 \AA. The lattice constant for the $3R$ WSe$_2$ bilayer turned out to be 3.402 \AA, yielding the following vectors among the nearest local minima (purple points) on Figure \ref{fig:Fig1}(b): $\mathbf{s}_1=(1.964,0.000,0.000)$ \AA, $\mathbf{s}_2=(-0.982,1.701,0.000)$ \AA, and $\mathbf{s}_3=(-0.982,-1.701,0.000)$ \AA, respectively, where the letter $s$ stands for {\em sliding}. These vectors are facilitated as an inset on Figure \ref{fig:Fig3}(e).

At 490 K (Figure \ref{fig:Fig3}(f)), numerical averages indicate a sudden, discrete sliding from one {\color{black}degenerate} minima onto a nearest one at location $\mathbf{s}_2$ {\color{black}(see Figure \ref{fig:Fig3}(e) for a definition of vectors $\mathbf{s}_1$, $\mathbf{s}_2$, and $\mathbf{s}_3$)}. Two atomistic snapshots, one taken at 0.1 $\mu$s and the other at 0.1 $\mu$s and displayed on Figure \ref{fig:Fig4}, verify a sliding event in which all atoms on a given monolayer {\em moved in unison}, while their side views confirm that a swap of polarization $P$ has taken place \cite{Franchini,2018paper,park2019ferroelectric,Liu,pablo}  {\em i.e.}, changing from $\mathbf{up}$ to $\mathbf{down}$. Figure  \ref{fig:Fig3} confirms the hypothesis that a paraelectric $3R$ TMDB is a time-average of swapping ferroelectric structures.

With increasing temperature (subplots \ref{fig:Fig3}(g) to \ref{fig:Fig3}(i)), one continues to see sudden jumps that continue to confirm that the bilayer is exploring the infinite number of minima freely, and we posit that $T_C$ is 490 K for the $3R$ WSe$_2$ bilayer. This value is higher than the experimentally reported one of 351 K \cite{Liu}, but it is order-of-magnitude correct. The discrepancy can be used to revise and tune the exchange-correlation potential employed for the training of the force field.

To the argument that the rare slippage events could occur at even lower temperatures if one continues tracking the temporal evolution for longer times, one must recall that there is an activation barrier $J_{3R}$ that must be overcome here, and that the relation among $T_C$ and $J$ for the $3R$ WSe$_2$ bilayer just found is consistent with previous results on other 2D ferroelectrics, which indicate a relation in between 1 and 2 among those two physical variables when $J$ is expressed in K/u.c. We show that every additional sliding event swaps $P$ on the SI.

{\color{black}Slippage events turn rarer as the supercell employed increase in size. The shear mode's probability diminishes when more atoms are used in simulations and prohibitive, larger than the microsecond times reported here, are needed to capture those events.} The paraelectricity of $3R$ WSe$_2$ bilayers occurs at a definite temperature experimentally\cite{Liu}, and we are positing that such observation may be {\em a time-average} of suddenly swapping ferroelectric configurations over long times. Experimental confirmation {\color{black}of our hypothesis} may come from time- and spatially-resolved ferroelectric probes \cite{sabine1,sabine2,sabine3,sabine4}. The relatively slow swapping time leading to the paraelectric state here is to be contrasted with the few nanoseconds it takes for group-IV monochalcogenide monolayers to turn paraelectric \cite{MehboudiNL}, which is a hundred times faster.

\begin{figure}
\begin{center}
\includegraphics[width=0.48\textwidth]{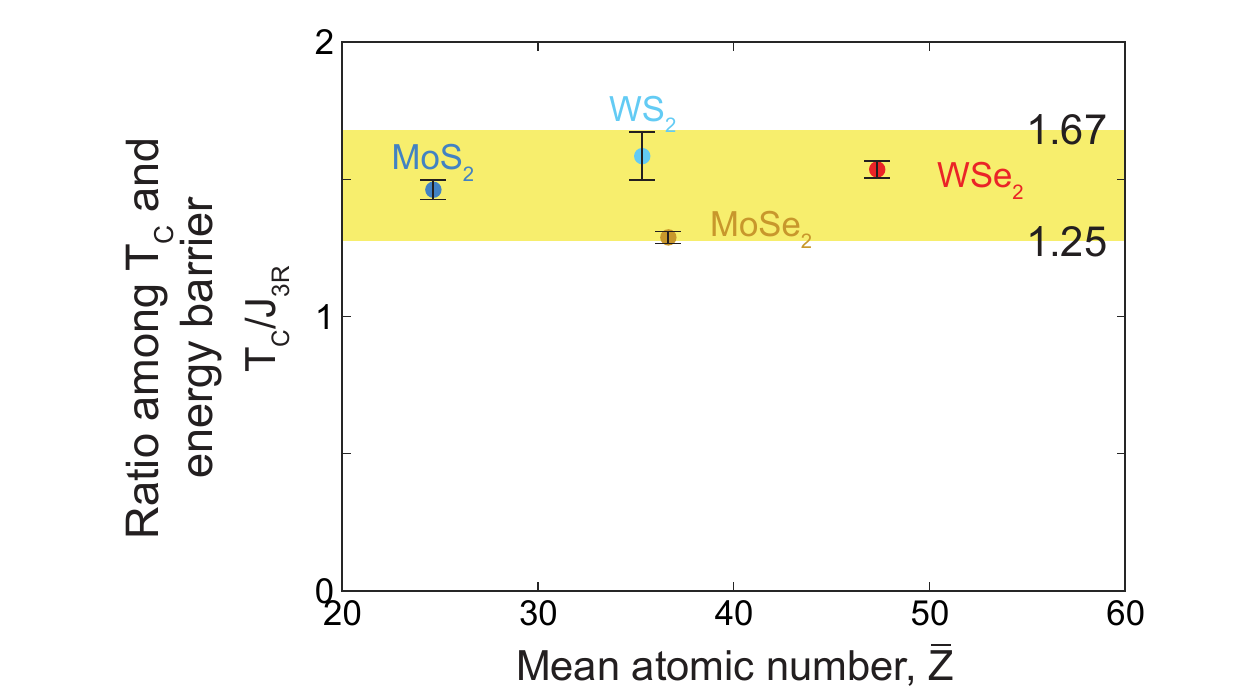}
\end{center}
\caption{The ratio among $T_C$ and $J_{3R}$ {\color{black}lies} in between 1.25 and 1.67. Vertical error bars account for the 10 K uncertainty on $T_C$. This ratio can be utilized to quickly estimate $T_C$ once the energy barrier $J_{3R}$ is known. \label{fig:Fig5}}
\end{figure}

This work ends with additional calculations to ascertain $T_C$ for WS$_2$, MoS$_2$, and MoSe$_2$ $3R$ bilayers following the procedure described on Figure \ref{fig:Fig3} (SI), and we obtained $T_C=180$, 410, and 590 K, for the WS$_2$, MoS$_2$ and the MoSe$_2$ bilayers, respectively. The ratio $T_C/J_{3R}$ is plotted on Figure \ref{fig:Fig5}, showing a relation among those variables consistent with previous finding for other 2D ferroelectrics {\color{black}($J_{3R}<T_C<2J_{3R}$)} \cite{prb2018}. Although specific values for $T_C$ as obtained here may differ from experimental estimates, the novel phenomenology thus described {\color{black}helps} make sense of the observed paraelectric behavior of these chemically inert and ultra-novel 2D ferroelectrics.

Proceeding by comparison with another family of 2D ferroelectrics, it has been shown that the ``paraelectric phase'' of $3R$ transition metal dichalcogenide bilayers is a time average over large times of a sequence of ferroelectric configurations that swap polarization sequentially over a {\em periodic} energy landscape. This conclusion is supported by a study of the energy landscape, vibrational modes, and dedicated molecular dynamics calculations. These results invite to rethink the atomistic and temporal nature of ferroelectrics made out of bilayers that slide easily.

\section{Methods}

\subsection{Zero-temperature calculations:} {\color{black}We} used the VASP package \cite{vasp-1,vasp-2,vasp-4} with projector augmented wave (PAW) pseudopotentials, and the opt-PBE GGA exchange-correlation functional to account for van der Waals forces \cite{Klime_2009,Klime_2011}, which was shown to provide accurate structural and energy barrier estimations for other 2D ferroelectrics \cite{shiva}. A Monkhorst-Pack mesh including $21\times21\times1$ $k$-points,  an energy convergence criterion of $10^{-8}$ eV, and a cutoff energy of $600$ eV were utilized. All calculations include dipole moment energy corrections along the direction perpendicular to the periodic lattice, {\color{black} and the out-of-plane lattice constant was set to}30 \AA.  Atomic positions and lattice vectors were relaxed {\color{black}down to} $10^{-2}$ eV/\AA.

To calculate energy landscapes, we performed a rigid shift of the top monolayer from $\mathbf{r}=\mathbf{0}$ to $\mathbf{r}=\mathbf{a}_1+\mathbf{a}_2$, with $\mathbf{a}_1$ and $\mathbf{a}_2$ lattice vectors, and the total energy was determined self-consistently for each structure along this translation. $P$ is calculated following the standard Berry-phase approach \cite{vanderbilt}.

{\em Ab initio} calculations {\color{black}to train} the classical force field (MTP) {\color{black}were performed with} the QuantumATK package \cite{atk,Smidstrup2020}, where the Kohn-Sham (KS) Hamiltonian is represented on a basis of double-zeta plus polarization (DZP) orbitals, using a density mesh cut-off of 105 Hartree, and a 4 \AA~ $k-point$ density along both $a_1$ and $a_2$. {\color{black}Exchange-correlation} interactions {\color{black}were} described with the Perdew-Burke-Ernzerhof (PBE) parametrization of the generalized gradient approximation (GGA) \cite{Perdew1996}, with semiempirical Grimme DFT-D2 dispersion corrections \cite{Grimme2006}. We minimized the volume and atomic coordinates with energy, force and stress criteria of 10$^{-3}$ eV, 10$^{-2}$ eV/\AA, and 0.1 GPa, respectively.

\subsection{Finite-temperature calculations:} We generated MTPs for MoS$_2$, MoSe$_2$, WS$_2$, and WSe$_2$ {\color{black}$3R$ bilayers}, for which we used fully relaxed hexagonal unit cells with lattice vectors $a_0=3.2274$, 3.3846, 3.2242 and 3.4025 \AA, respectively, and replicated those to create initial 5$\times$5$\times$1 supercells. Each training set consisted of 142 system configurations, from which 42 are obtained with molecular dynamics on the NPT ensemble with zero target pressure during 200 fs, after an NVT temperature equilibration during 200 fs using the Nos{\'e}-Hoover thermostat and a 1 fs time step. Temperature is set to 500 K, and snapshots are taken every 10 fs. The remaining configurations are obtained from small random displacements to the atomic coordinates in the supercell with up to 0.15 \AA~ atomic rattling amplitude. Expanding/contracting the lattice vector within $\pm$5\% to 20\% {\color{black}gave us additional structures for testing energetics}.

\section{Author Contributions}
SBL and MAM conceived the project. JMMT trained the machine-learning interatomic potentials for the molecular dynamics calculations, and ran those to find the sliding events that underpin the paraelectric transformation. Along with SPP, JMMT developed energy landscapes by sliding of the $3R$ phase. JER calculated one-dimensional cuts of the energy landscape for the $2H$ and $3R$ bilayers and determined the energy barriers to overcome. JER computed the intrinsic dipole moments with aid from SPP. JMMT and SPP calculated phonon dispersions and lowest energy eigenvectors and energy landscapes. All authors discussed the results. SBL wrote the manuscript with input from all authors.

\section{Notes}
The authors declare no competing financial interest.

\begin{acknowledgement}

JMMT and MAM thank Montana State University, Bozeman, for startup support and computational
resources within the Tempest Research Cluster. Calculations from the Arkansas team were performed at Cori at NERSC, a DOE facility funded under contract No. DE-AC02-05CH11231, and at the University of Arkansas’ Pinnacle supercomputer, funded by the U.S. National Science Foundation, the Arkansas Economic Development Commission, and the Office of the Vice Provost for Research and Innovation. All authors acknowledge financial support from the MonArk NSF Quantum Foundry, supported by the National Science Foundation Q-AMASE-i program under NSF award No. DMR-1906383. {\color{black}Conversations with P.~Kumar and A. Pacheco San Juan are gratefully acknowledged.}

\end{acknowledgement}

\begin{suppinfo}

Supporting Information includes structural models of $2H$ and $3R$ bilayers, additional force field fitting and benchmarking information, phonon dispersion calculations, evidence for additional swapping, and trajectories underpinning $T_C$ for additional compounds.

\end{suppinfo}

\begin{table}[tb]
\caption{ {\color{black}Energy cost} $\Delta E$ {\color{black}to turn the} ground-state $2H$ {\color{black}bilayers onto the} $3R$ {\color{black}ones}. $\Delta E$ increases with $\bar{Z}$.\label{ta:Ta1}}
\begin{tabular}{c|cc||c|cc}
\hline
\hline
Chemical &$\bar{Z}$ & $\Delta E$         & Chemical &$\bar{Z}$ & $\Delta E$  \\
formula  &          & (K/u.c.)           & formula  &          & (K/u.c.)\\
\hline
MoS$_2$	 &24.667	&  $+$5.895 &  WS$_2$	 &35.333	& $+$19.055\\
MoSe$_2$ &36.667	& $+$21.619 & WSe$_2$	 &47.333	& $+$38.504\\
\hline
\hline
\end{tabular}
\end{table}

\newpage

\providecommand{\latin}[1]{#1}
\makeatletter
\providecommand{\doi}
  {\begingroup\let\do\@makeother\dospecials
  \catcode`\{=1 \catcode`\}=2 \doi@aux}
\providecommand{\doi@aux}[1]{\endgroup\texttt{#1}}
\makeatother
\providecommand*\mcitethebibliography{\thebibliography}
\csname @ifundefined\endcsname{endmcitethebibliography}
  {\let\endmcitethebibliography\endthebibliography}{}

\end{document}